\documentclass{ws-ijmpa}
\usepackage{graphicx}
\newcommand{\SLASH}[1]{/\!\!\! #1}

\begin{document}

\markboth{B.~El-Bennich, O.~M.~A.~Leitner, B.~Loiseau, J.-P.~Dedonder}{Form factors in $B\to f_0(980)$ and $D\to f_0(980)$ 
                   transitions from dispersion relations}
                   
%
\catchline{}{}{}{}{}
%

\title{FORM FACTORS IN $B\to f_0(980)$ AND $D\to f_0(980)$ TRANSITIONS FROM DISPERSION RELATIONS}

\author{B.~El-BENNICH$^1$, O.~M.~A.~LEITNER$^1$, B.~LOISEAU$^1$ and J.-P.~DEDONDER$^2$}
\address{$^1$ Laboratoire de Physique Nucl\'eaire et de Hautes \'Energies,  Groupe Th\'eorie \\
                 Universit\'e Pierre et Marie Curie, 4 place Jussieu, 75252 Paris, France\\ bruno.elbennich@lpnhe.in2p3.fr}
\address{$^2$ GMPIB, Universit\'e Denis Diderot, case 7021-- F-75251 Paris, France}

\maketitle


\begin{abstract}
Within the dispersion relation approach we give the double spectral representation for space-like and time-like $B\to f_0(980)$ and 
$D\to f_0(980)$ transition form factors in the full $q^2$ range. The spectral densities, being the input of the dispersion relations, 
are obtained from a triangle diagram in the relativistic constituent quark model.
\keywords{Transition form factors; bottom and charm mesons; relativistic quark model}
\end{abstract}

\ccode{PACS numbers:11.55.Fv, 12.39.Ki, 13.20.He, 13.25.Ft}

\section{Scalar mesons in heavy-meson decays}	

The structure of scalar mesons in terms of fundamental constituents is still an open question and widely debatted, for a review see Refs.~\refcite{review}. 
We are interested in the scalar-isoscalar meson $f_0(980)$, but the formalism we are going to present in this talk is of general interest and applicable 
to any scalar meson. Besides the quantum numbers, what is known about the $f_0(980)$ is its mass and width: $m=980\pm 10$~MeV and 
$\Gamma=70\pm 30$~MeV. Hence, the $f_0(980)$ is a rather narrow state, already known, as well as other scalar resonances, for over thirty years
from $\pi\pi$ scattering.\cite{Protopopescu:1973sh} A general analysis of data leads to a picture suggesting that the scalar mesons above 1~GeV 
can be identified as conventional $\bar qq$ nonet with some possible admixture of gluons, whereas light scalar mesons below or near 1~GeV form 
a flavor nonet {\em predominantly} made of ${\bar q}^2 q^2$ states.\cite{Krupa:1995fc} The  $f_0(980)$ being the product of a resonant 
two-pion state at the hadronic level may lead one to think of it as a four-quark state, which is also favored by the observation that such a configuration 
yields the right quantum numbers without the need for an orbital angular momentum to make a $0^+$ state. Moreover, studies of the scalar mass 
spectrum below 1~GeV suggest a ${\bar q}^2 q^2$ state as originally advocated by Jaffe.\cite{Jaffe} On the other hand, scalar mesons produced
in $D$ and $D_s$ decays leading to three pseudoscalar mesons\cite{focus} seem to indicate a $\bar qq$ content, although higher statistics in these
experiments is desirable.  

The emergence of the $f_0(980)$ as a pole of the $\pi\pi$ amplitude in the $S$-wave\cite{Kaminski:1998ns} is also well established in three-body 
decays of $B$-mesons. Recently, distinct peaks about 1~GeV were observed in the $\pi\pi$ effective mass range distributions of $B\to f_0(980)K$ 
decays by the BaBar and Belle Collaborations.\cite{massdistrib} As of yet, these newer data do not allow to discriminate between a two-quark or a 
four-quark description of the $f_0(980)$.  Nonetheless, viewing the $f_0(980)$  {\em exclusively} as a $\bar qq$ or a ${\bar q}^2 q^2$ state may prove 
to be misleading. In the case of $B\to f_0(980)K$ decays there are plausible reasons to limit oneself to the $\bar qq$ picture of the naive quark model. 
Due to the large $B$-mass, the outgoing mesons behave as massless particles, which prompts to expand the corresponding bound states in terms 
of Fock states. Quark configurations  like ${\bar q}^2 q^2$ or ${\bar q}^2 q^2 g$ therefore belong to higher Fock states. It was already suggested by 
Cheng, Chua and Yang\cite{cheng} that the $\bar qq$ component of the energetic $f_0(980)$ may be more important, as {\em two} rapid $\bar qq$ 
pairs are less likely to form a fast moving $f_0(980)$. In our model we therefore neglect higher Fock contributions to the bound state.

\section{Transition form factors}

After we have settled down to the $\bar qq$ state of the $f_0(980)$, we ought to be concerned with its flavor content. Since the branching
ratios $\mathcal{B}(J/\Psi \to f_0(980)\phi )$ and $\mathcal{B}(J/\Psi \to f_0(980)\omega )$ have almost the same magnitude, the $f_0(980)$
state must necessarily contain $\bar uu$, $\bar dd$ as well as $\bar ss$ components. We write the wave function as
\begin{equation}
   \Psi_{f_0} = \frac{1}{\sqrt{2}} \big ( |\bar uu \rangle + |\bar dd \rangle \big )\sin \theta_m + |\bar ss \rangle \cos\theta_m = 
   \Phi^n \sin\theta_m +\Phi^s \cos \theta_m,
\label{wavefct}
\end {equation}
where $\theta_m$ is the mixing angle and $n=u, d$. 

\begin{figure}[t]
\vspace*{1.6cm}
\begin{center}
\includegraphics[scale=0.5]{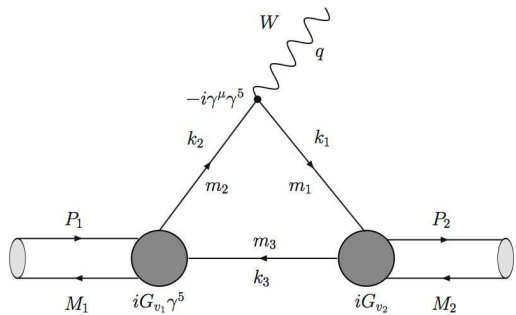}
\caption{The triangle diagram describing the weak-hadronic $P\to S$ transition amplitude in relativistic constituent quark model.}
\label{fig1}
\end{center}
\end{figure}

We are concerned with an effective description of hadronic corrections to the weak vertex in charged current-quark transitions $b\to u$, $c\to s$ which 
drive the weak decays $B\to f_0^{u,s}(980)$ or $D\to f_0^{u,s}(980)$. The exponents $u, s$ denote the flavor content of the scalar $\bar qq$ state. 
To this end, we calculate the pseudoscalar to scalar $(P\to S)$ transition amplitude $\langle S(p_2) |J^\mu |P(p_1)\rangle$ corresponding to the 
diagram in Fig.~\ref{fig1} in a relativistic quark model. The momenta $p_1^2=M_1^2$ and $p_2^2=M_2^2$ belong to the initial pseudoscalar and 
final scalar meson, respectively. The transition amplitude is generally decomposed into two terms
\begin{equation}
  \langle S(p_2) | J^\mu | P(p_1)\rangle = F_+(q^2)\, (p_1+p_2)^\mu  + F_- (q^2)\, (p_1-p_2)^\mu ,
\end{equation}
where $F_+(q^2)$ and $F_-(q^2)$ are the transition form factors, $q=p_1-p_2$ and the weak current is $J^\mu = \bar q' \gamma^\mu (1-\gamma^5) q$ 
with $q'=u,d,s$ and $q=c,b$.

In the dispersion relation approach, the form factors are derived from the diagram in Fig.~\ref{fig1} and expressed through the 
double spectral representation as
\begin{equation}
 F_\pm (q^2) = \int \frac{ds_1 G_{v_1}(s_1)}{\pi (s_1-M_1^2) } \frac{ds_1 G_{v_2}(s_2)}{\pi (s_2-M_2^2)} \,\Delta_\pm (s_1,s_2,q^2; m_1,m_2,m_3).
\label{dispersion}
\end{equation}
The vertex $G_v(s)$ describes the bound-state transition to the constituent quarks and $\Delta_\pm$ are spectral densities to be discussed
below. The wavefunction of a bound state with negative energy is related to the vertex by $ \Phi (s) = G_v (s)/(s-M^2)$ 
where, at the pole, $M^2$ is the mass of the bound state. In the case of $\bar qq$ pairs, nevertheless, the strong interaction properties lead
to confinement, commonly approximated by a harmonic oscillator, and the bound state occurs at $s=M^2>(m_q+m_{\bar q})^2$. Following
the diagram in Fig.~\ref{fig1}, $m_q$ and $m_{\bar q}$ denote the mass pairs $(m_1+ m_3)^2< M_2^2$ and $(m_2+ m_3)^2 <M_1^2$ for the 
scalar and pseudoscalar meson, respectively. Confinement does smear out the pole and the smooth form of the wavefunction is the usual 
Gaussian one. It is therefore more convenient to use the wavefunction $\Phi (s)$ in Eq.~\eqref{dispersion} rather than the vertex function, as 
the pole position cancels. The normalization of the vertices $G_v(s)$ and therefore of $\Phi (s)$  is determined by the rescattering of the two 
constituent quarks in the meson. Including the proper relativistic normalization, a wavefunction like that in Eq.~\eqref{wavefct} is given by
\begin{equation}
  \Phi (s) =  \xi_\pm (s)\, \exp ( -4 \alpha k^2/ \mu^2) \, ,
\label{Phi}
\end{equation}
where $\mu=m_q m_{\bar q}/(m_q+m_{\bar q})$ is the reduced mass of the $\bar qq$ pair, $k$ is the modulus of the center-of-mass quark
momentum and\cite{melikhov}
\begin{equation}
  \xi_\pm (s) = \sqrt{2}  \left ( \frac{8\pi \alpha }{s \mu^2} \right )^{\!\!3/4} \!
                         \sqrt{\frac{s^2 - (m_q^2-m_{\bar q}^2)^2}{s-(m_q\pm m_{\bar q})^2}}  . 
\end{equation}
In Eq.~\eqref{Phi}, $\xi_+(s)$ is the normalization of a scalar meson $\Phi_S(s)$ whereas $\xi_-(s)$ is that of a pseudoscalar meson $\Phi_P(s)$.
The parameter $\alpha$ controls the size of the meson and has to be deduced phenomenologically. This can be done for $B$- and $D$-mesons 
by using constraints like the meson decay constant known from experiment. 
For neutral scalar mesons the situation is more tricky --- owing to charge conjugation invariance, the $f_0(980)$ cannot be produced via a vector 
current. Alternatively, one defines a scalar decay constant $f_S$ associated with the scalar current, although its value is only vaguely known 
from theoretical estimations. We prefer to determine $\Phi^{u,d}_S(s)$ and $\Phi^s_S(s)$ by fitting $D(D_s)\to f_0(980) X$ $(X=\pi, K)$ branching 
fractions, as will be discussed in more detail in Ref.~\refcite{bennich}.

\begin{figure}[t]
\begin{center}
\includegraphics[scale=0.8]{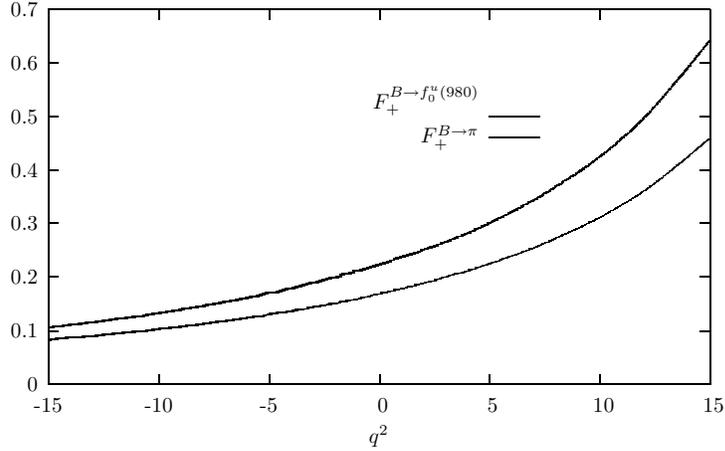}
\caption{The form factor $F^+(q^2)$ for the $P\to S$ transition $B\to f_0^u(980)$ (lower curve) compared to the $P\to P$ transition $B\to \pi$ (upper curve).}
\label{fig2}
\end{center}
\end{figure}

The spectral densities $\Delta_\pm$ in Eq.~\eqref{dispersion} are obtained from the double discontinuities of the triangle diagram.\cite{melikhov} 
We apply the Landau-Cutkosky rules, thus the internal quarks are put on-mass shell ($k_i^2=m_i^2,\, i=1,2,3)$, to calculate these discontinuities for 
a $P\to S$ transition:
\begin{eqnarray}
\lefteqn{(\tilde p_1+\tilde p_2)^\mu\Delta_+ (s_1,s_2,q^2;m_1,m_2,m_3)+ (\tilde p_1 - \tilde p_2)^\mu\Delta_- (s_1,s_2,q^2;m_1,m_2,m_3) =} 
\nonumber \\
&=& \frac{1}{8\pi}\int\! dk_1dk_2 dk_3\,\delta(k_1^2-m_1^2)\delta(k_2^2-m_2^2) \delta(k_3^2-m_3^2) \delta(\tilde p_1-k_2-k_3)\delta(\tilde p_2-k_3-k_1) 
\nonumber \\ 
& &\hspace{2cm} \times\, \mathrm{Tr}\left [-(\SLASH k_{1}+m_1)\gamma^\mu\gamma^5(\SLASH k_{2}+m_2)i\gamma^5 (m_3-\SLASH k_{3})i \right ].  
\end{eqnarray}
One has the condition $m_2 > m_1$ and the tilda on the external momenta expresses their off-shellness in the dispersion approach with 
$s_1=\tilde p_1^2$ and $s_2=\tilde p_2^2$. These momenta are the dynamical variables in the double dispersion relation of Eq.~\eqref{dispersion}.

\section{Numerical applications}

Having the explicit form of the spectral densities, one can calculate the space-like form factors $F_+(q^2)$ and $F_-(q^2)$ for $q^2<0$ and an analytical
continuation in $q^2$ yields the time-like equivalent for $0<q^2<(M_1-M_2)^2$. Thus, the dispersion relations applied to the relativistic quark model 
allow to derive $B\to f_0$ and $D\to f_0$ transition form factors for the full momentum range, and in particular for the kaon mass $q^2=m_K^2$
needed in the decays $B\to f_0(980)K$. This is done without any extrapolation and the quark model solely determines the behavior of the form
factors. As an example, we show the {\em preliminary} function $F_+(q^2)$ for the $P$ to $S$ transition $B\to f_0^u(980)$ compared to that for the
$P$ to $P$ transition $B\to \pi$ in Fig.~\ref{fig2}. In an upcoming paper,\cite{bennich} we will give an estimate for the errors of these transition form 
factors. With this work, we hope to better determine one component of the hadronic matrix elements in the factorization approach for $B$- and 
$D$-decays into scalar mesons, for which most of the uncertainties stem from the non-perturbative character of QCD.

\end{document}